\documentclass[10pt,superscriptaddress]{revtex4-2}
\usepackage[colorlinks,allcolors=blue]{hyperref}
\usepackage{amsmath,amsfonts,tikz}
\usepackage{graphicx,xcolor,multirow}

\usepackage{lineno}
\usepackage{natbib}
\begin{document}
%\linenumbers
\title{Quantum optimization within lattice gauge theory model on a quantum simulator}
%\title{Quantum annealing meets topology}
\author{Zheng Yan}
%\thanks{These authors contributed equally to this work.}
\email{zhengyan@westlake.edu.cn}
\affiliation{Department of Physics, School of Science, Westlake University, Hangzhou 310030, China}
\affiliation{Institute of Natural Sciences, Westlake Institute for Advanced Study, Hangzhou 310024, China}
%\affiliation{Department of Physics and HKU-UCAS Joint Institute of Theoretical and Computational Physics, The University of Hong Kong, Pokfulam Road, Hong Kong, China}
%\affiliation{State Key Laboratory of Surface Physics and Department of Physics, Fudan University, Shanghai 200438, China}

\author{Zheng Zhou}
%\thanks{These authors contributed equally to this work.}
%\affiliation{State Key Laboratory of Surface Physics and Department of Physics, Fudan University, Shanghai 200438, China}
\affiliation{Perimeter Institute for Theoretical Physics, Waterloo, Ontario N2L 2Y5, Canada}
\affiliation{Department of Physics and Astronomy, University of Waterloo, Waterloo, Ontario, Canada N2L 3G1}

\author{Yan-Hua Zhou}
\affiliation{Department of Physics, and Center of Quantum Materials and Devices, Chongqing University, Chongqing 401331, China}
\affiliation{Chongqing Key Laboratory for Strongly Coupled Physics, Chongqing University, Chongqing 401331, China}

\author{Yan-Cheng Wang}
\affiliation{Zhongfa Aviation Institute of Beihang University, Hangzhou  311115, China}
\affiliation{Tianmushan Laboratory, Hangzhou 310023, China}

\author{Xingze Qiu}
\email{xingze@fudan.edu.cn}
\affiliation{State Key Laboratory of Surface Physics and Department of Physics, Fudan University, Shanghai 200438, China}

\author{Zi Yang Meng}
\email{zymeng@hku.hk}
\affiliation{Department of Physics and HKU-UCAS Joint Institute of Theoretical and Computational Physics,
The University of Hong Kong, Pokfulam Road, Hong Kong, China}

\author{Xue-Feng Zhang}
\email{zhangxf@cqu.edu.cn}
\affiliation{Department of Physics, and Center of Quantum Materials and Devices, Chongqing University, Chongqing 401331, China}
\affiliation{Chongqing Key Laboratory for Strongly Coupled Physics, Chongqing University, Chongqing 401331, China}

\begin{abstract}
\noindent{\bf Abstract} Simulating lattice gauge theory (LGT) Hamiltonian and its nontrivial states by programmable quantum devices has attracted numerous attention in recent years. Rydberg atom arrays constitute one of the most rapidly developing arenas for quantum simulation and quantum computing. The $\mathbb{Z}_2$ LGT and topological order has been realized in experiments while the $U(1)$ LGT is being worked hard on the way. States of LGT have local constraint and are fragmented into several winding sectors with topological protection. It is therefore difficult to reach the ground state in target sector for experiments,
and it is also an important task for quantum topological memory. Here, we propose a protocol of sweeping quantum annealing (SQA) for searching the ground state among topological sectors. With the quantum Monte Carlo method, we show that this SQA has linear time complexity of size with applications to the antiferromagnetic transverse field Ising model, which has emergent $U(1)$ gauge fields. This SQA protocol can be realized easily on quantum simulation platforms such as Rydberg array and D-wave annealer. We expect this approach would provide an efficient recipe for resolving the topological hindrances in quantum optimization and the preparation of quantum topological state.
\end{abstract}

\maketitle

%The advantages of quantum annealer
\section{Introduction}
Rydberg atom arrays constitute one of the most rapidly developing arenas for quantum simulation and quantum computing. They offer Ising-like interactions between qubits and single site manipulation, and also the ability to arrange hundreds of qubits in arbitrary geometry. This also opens up remarkable opportunities for studying topological phases of matter with fractional excitations.
Recent quantum simulation advances have provided remarkable microscopic access to the quantum correlations of a $\mathbb{Z}_2$ quantum spin liquid (QSL) \cite{Roushan21,Semeghini21}.
The $\mathbb{Z}_2$ QSL \cite{RS91,wen1991} is the simplest quantum state in two spatial dimensions with fractionalized excitations and time-reversal symmetry, and has the same anyon content as the toric code \cite{Kitaev1997}. One of the most interesting directions in quantum simulation is to construct lattice gauge theory (LGT) Hamiltonian and its nontrivial quantum states via Rydberg arrays, superconducting circuits and other platforms~\cite{Roushan21,Semeghini21,PhysRevLett.118.080502,PRXQuantum.3.020320,yan2023glass}. Quantum dimer model as a typical LGT is widely prepared in certain highly frustrated parameter region of Rydberg arrays~\cite{RK1988,YanZheng2019b,yan2020triangular,QDMbook,yan2022triangular,ZYloop2022,zhou2022Rydberg,ran2023fully}. LGT models have Hilbert space fragmentation due to the local constraints, the sub-Hilbert spaces are labelled by different winding numbers and protected by topological defects. The topological defects also induce nontrivial phenomena, such as incommensurate phases and Cantor deconfinement~\cite{PhysRevB.69.224415,PhysRevLett.115.217202}. 
For example, in the incommensurate phase, each topological sector becomes the ground state in turn while tuning the parameter. In the thermodynamic limit, it means infinite topological sectors one by one arise as the ground state in a finite region of parameter, which is also called `devil stairs'~\cite{PhysRevLett.42.122,bak1982commensurate}. But it is a hardcore problem in experiment to make the state of system changed from one sector to another while tuning related parameters, because the topology always hinders the jumping between different sectors.
At present, there is no control scheme of experiment that can overcome the topological protection and prepare/search quantum states in a certain topological sector accurately.
%Once we include considerations of lattice and other symmetries,  $\mathbb{Z}_2$ QSLs come in different varieties; the distinctions between them are important in understanding the phase diagrams of possible experimental realizations. The coarsest classification subdivides $\mathbb{Z}_2$ QSLs into `odd' and `even' classes, depending upon whether elementary translations anti-commute or commute when acting on excitations carrying $\mathbb{Z}_2$ magnetic flux \cite{RJSS91,MVSS99,TSMPAF99,Moessner01}, and results in different translational symmetry fractionalization patterns and spectral signatures in the dynamic response \cite{Essin2014,JWMei2015,YCWang2018,YCWang2017QSL,YCWang2021NC,GYSun2018}. More refined classifications have been obtained since \cite{Essin:2013rca,Zaletel:2014epa,Cheng:2015kce,QiMeng16,Bulmash:2020flp}.
\begin{figure}[htp]
    \includegraphics[width=\columnwidth]{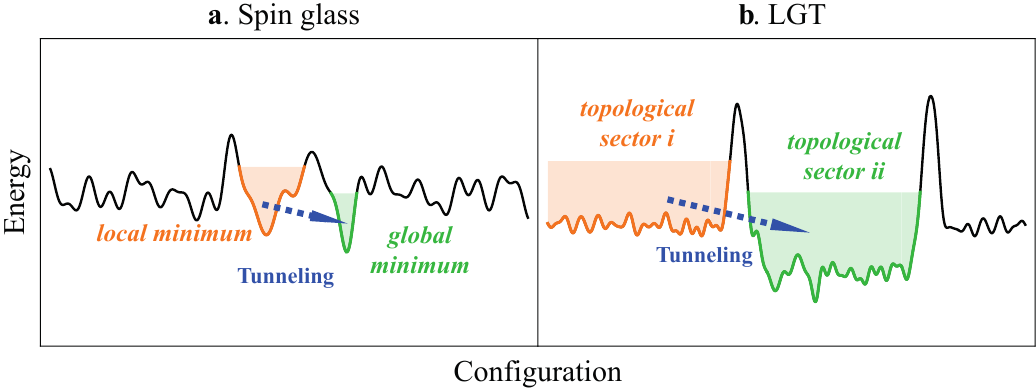}
    \caption{
    \textbf{Energy-Configuration landscape.} Conventional optimization problems in glassy systems (a) and topological optimization problems in lattice gauge theory (LGT) cases (b).
    }
    \label{Fig1}
\end{figure}

On the other hand, quantum annealing (QA) method is a powerful tool for optimization of Ising-like encoded Hamiltonian~\cite{Anderson1986,Virasoro1987,Huse1986,Car2002,Lucas2014,kadowaki1998quantum,Car2002,Montanari2009,Heim2015}, which utilizes the quantum fluctuation to approach the ground state. The degree of quantum acceleration depends very much on the design of QA algorithm~\cite{somma2012quantum,das2005quantum,Das2008,ebadi2022quantum},
especially for the lattice gauge model \cite{Lukin2022Dynamical}.
Following recent technological advancements in manufacturing coupled qubit systems, the QA algorithm can be embedded into superconducting flux qubits~\cite{Gildert2011,Boixo2013,Boixo2014}. Currently, QA computers, i.e., quantum annealer have also been commercialized, such as the D-wave machine, and it does show higher efficiency than classical computers in certain optimization problems~\cite{perdomo2012finding,novotny2016spanning,qiu2020precise,qiu2020programmable,king2023quantum}. Recently, Rydberg array simulator can even realize the Ising-like encoded Hamiltonian in large scale with high tunability~\cite{Roushan21,Semeghini21}, which is thus an ideal QA implementation platform.

The normal optimization problems usually find the ground states of glassy Hamiltonian as Fig.~\ref{Fig1} (a). In this field, the technology has been developed very well and matured~\cite{Anderson1986,Virasoro1987,Huse1986,Car2002,Lucas2014,kadowaki1998quantum,Car2002,Montanari2009,Heim2015,King2021PRXQuantum}. {As the fast development of quantum control and simulation, especially recent Rydberg arrays experiments, the optimization problem of LGT Hamiltonian takes another hardcore problems of simulation [Fig.~\ref{Fig1} (b)], that is, how to reach the target topological sector containing the ground state. }
As mentioned in the introduction, for instance, the simulation of `devil stairs' phenomenon requires the system can approach the ground states within different sectors under different parameters.
%In the past, much attention has been paid to the glassy case with the energy landscape shown in Fig.~\ref{Fig1} (a), while the frustrated one in Fig.~\ref{Fig1} (b) have been overlooked.
As the complex topological defects emerged in LGT systems are robust to quantum fluctuations, it is realized that these emergent topological structures could deeply hinder the efficiency of the existing annealing algorithms. Therefore it is the time of finding useful quantum annealing schemes for such systems, hereby we give a powerful scheme, sweeping quantum annealing (SQA), to solve the problems. Different from the normal QA with local tunneling on each site, the SQA removes the topological defects by additional global annealing on virtual edges to overcome the barrier between the topological sectors. We demonstrate the effectiveness of SQA through some extremely egregious examples which can be realized in Rydberg experiments.

\begin{figure}[htp]
    \includegraphics[width=\columnwidth]{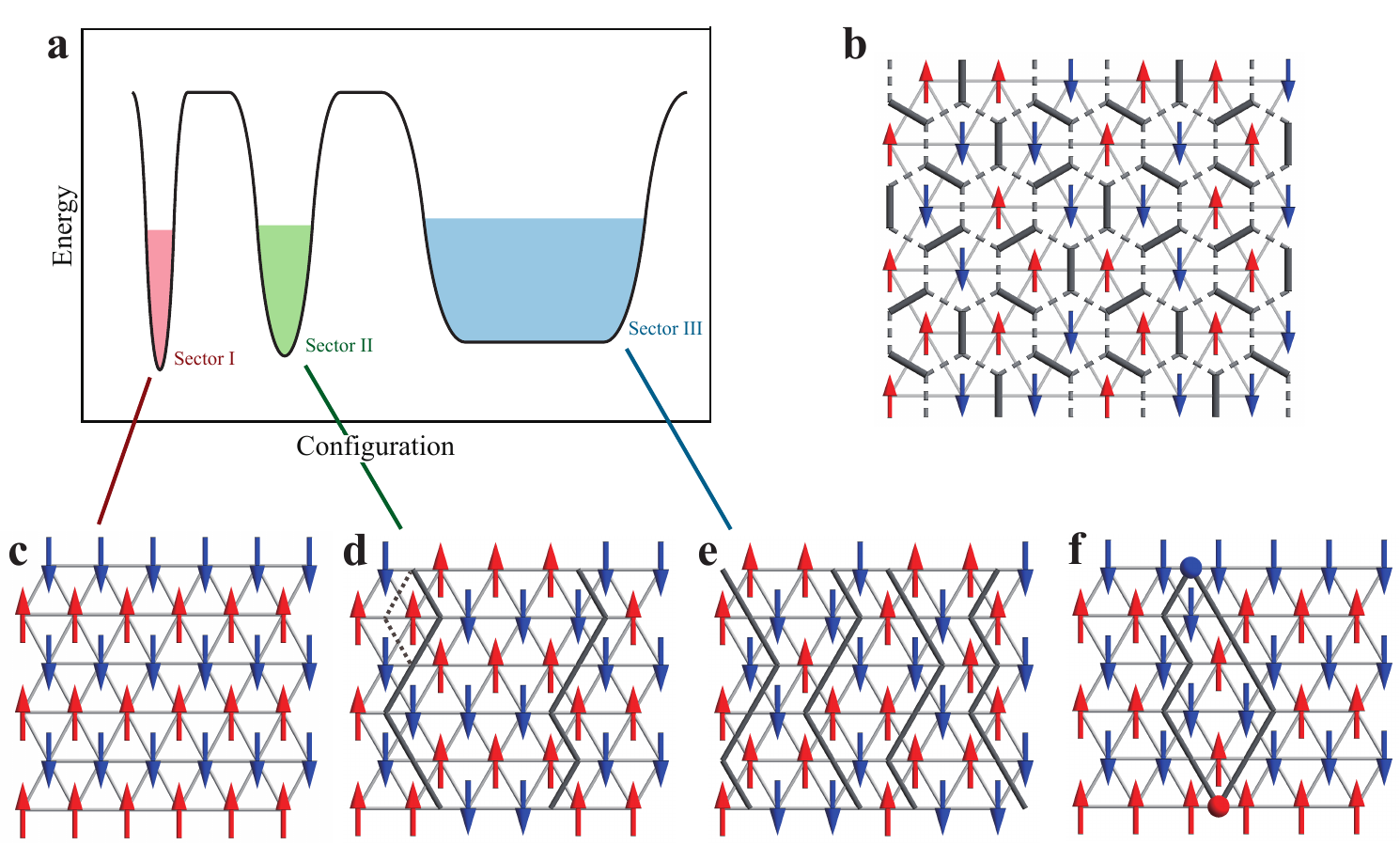}
    \caption{\textbf{The emergent topology in the triangular lattice AFM Ising model} (a) Schematic diagram for energy configurations of different topological sectors (with the three hardcore characteristics discussed in the main text). (b) The mapping between the constrained spin configuration and the dimers. (c) The stripe phase with no defect is the ground state of Hamiltonian Eq.~\eqref{HA} in $N_{\rm D}=0$ topological sector. (d) and (e) are the topological sector with $N_{\rm D}=2$ and $N_{\rm D}=4$ and the topological defects are denoted by the black lines. As shown in (d), spin on corner can be flipped without breaking the 'triangle rule'. The defect before/after spin flipped is labeled by solid/dashed line. (f) A pair of spinons (triangle with 3 parallel spins) connected with defects. The spinons can be considered as local defects and they can annihilate (with local flipping) when meet.}
    \label{Fig2}
\end{figure}

\section{Results}
\subsection{Model}
Without losing generality, we use U(1) LGT as an example to test the efficiency of the different QA algorithms. Because 2+1d $U(1)$ LGT has about $L^2$ topological sectors, more than $\mathbb{Z}_2$ LGT which has only four sectors~\cite{QDMbook,yan2020improved}. Moreover, we would design a hard-mode case to demonstrate the power of SQA. The difficult topological optimization problem has three characteristics as shown in (a) of Fig.\ref{Fig2}: 1) The minimum energies of many topological sectors are nearly equal. 2) The target topological sector containing the ground state occupies small Hilbert subspace, it is difficult to be found. 3) Larger topological sectors provide enough degree of freedom for quantum fluctuations in the annealing process to compete with the ground energy sector. In other words, annealing tends to pull the system into a wrong sector for a kinetic energy advantage. Therefore, we consider such a frustrated antiferromagnetic (AF) Ising model with small transverse field on a triangular lattice, whose Hamiltonian reads
\begin{equation}
    H_{\textrm{A}}=J_x\sum_{\langle ij\rangle_{x}}\sigma^z_i\sigma^z_j+J_{\wedge}\sum_{\langle ij\rangle_{\wedge}}\sigma^z_i\sigma^z_j-\delta\sum_i \sigma^x_i,
\label{HA}
\end{equation}
where $\sigma^z$ and $\sigma^x$ are the Pauli operator, ${\langle ij\rangle_{x}}$ and ${\langle ij\rangle_{\wedge}}$ represent the nearest-neighbor sites on the horizontal bonds and the interchain bonds, respectively, and $J_x$ and $J_{\wedge}$ are the corresponding coupling strength. The emergent LGT in this model requires the $\delta \ll J_x,J_{\wedge}$.
The emergent $U(1)$ gauge fields and topological properties in this model have been well-studied \cite{Moessner2001,Isakov,YCWang2017,YDLiao2021,zhou2020quantum}.
Due to the antiferromagnetic interaction, every triangle must be composed of two parallel and one antiparallel spins in the low-energy Hilbert space. We dub this local constraint as `triangle rule'', and the constraint-satisfying Hilbert space can be exactly mapped to a familiar quantum dimer model \cite{Moessner2010b, yan2019widely, ZY2019, zhou2020quantumstring,yan2020triangular, yan2020improved}.  Fig.~\ref{Fig2} (b) shows this mapping between the constrained spin configuration on triangular lattice and the dimer configuration on the dual honeycomb lattice, where the bond with two parallel spins corresponds to a dimer. The dimer density on the honeycomb lattice can be understood as lattice electric field on the dual bond, and the local constraint can be written as the divergenceless condition. There thus emerges an $U(1)$ gauge field in this triangular AF Ising model~\cite{Moessner2001,zhou2020quantum,ZYhqdm2022}, and the many-body configurations with constraints can be mapped to lattice electromagnetic fields which are naturally classified into different topological sectors~\cite{Moessner2001}. Recent numerical works~\cite{zhou2020quantum,ZZrk2021}, which studied a triangular lattice AFM Ising model with small transverse field ($\delta \ll J$), have demonstrated that the parameter range of Eq.(\ref{HA}) is very close to the famous Rokhsar-Kivelson point and incommensurate phase with Cantor deconfinement of a general quantum dimer model on honeycomb lattice. To demonstrate the power of our algorithm, we put the Hamiltonian on a periodic boundary lattice, in which case the topological defects are much more robust.

Topological sectors are robust to quantum fluctuations emerged in the LGT systems and can be labeled by the number of topological defects $N_{\rm D}$~\cite{zhang01,zhang02,zhang03,zhou2020quantum}. As examples, we show the spin configurations in different topological sectors in Fig.~\ref{Fig2} (c), (d) and (e) for $N_{\rm D}=0,\,2,\,4$, respectively, if we set the stripe configurations [Fig. ~\ref{Fig2} (c)] as the $N_{\rm D}=0$ reference state.
Note that all these topological sectors are degenerate when $J_x=J_{\wedge}$ and $\delta=0$, we thus set $J_x=0.9J_{\wedge}$ and $\delta\ll J_{\wedge}-J_x$ in the following to break the degeneracy weakly and make the stripe configuration [Fig.~\ref{Fig2} (c)] with $N_{\rm D}=0$ has lowest energy. This setting is for satisfying the hardcore three characteristics mentioned above, as shown in (a) of Fig.\ref{Fig2}. In fact, we can also set different configuration with target $N_{\rm D}$ as ground state via changing the related couplings $J_i$ of bond $i$.

The ground state stripe phase in Fig.~\ref{Fig2} (c) is composed of horizontal bonds connecting parallel spins and interchain bonds connecting antiparallel spins. Without breaking the local constraint of the `triangle rule', the number of horizontal antiparallel bonds is conserved in each row, and linking such antiparallel bonds on different chains will construct the one dimensional global (topological) defect [Fig.~\ref{Fig2} (d) and (e)].
Note that there can be a mass of spin configurations for the same defect number, and the topological sectors thus have extensive nearly degenerate quantum states.
In fact, the degeneracy is because flipping the spins at the corners of defects obeys the `triangle rule' with little energy cost, as shown in Fig.~\ref{Fig2} (d). 
Therefore, the degeneracy (degree of freedom) increases exponentially
with the defect number $N_{\rm D}$~\cite{QDMbook,zhou2020quantum,zhou2020quantumstring,yan2020improved}. Changing the defect number requires generating a pair of spinons
(constraint-breaking triangles) connected by defects [Fig.~\ref{Fig2} (f)], letting them go around the connected boundary to meet and annihilate. Such process changes $N_{\rm D}$ by two. However, exciting spinons and pulling them apart is extremely hard due to the large energy cost and global topology constraints, so that it is difficult to achieve the ground state energy sector from higher energy ones in this way.

It's worthy noting that, in the triangle rule limit ($\delta \ll J_x, J_{\wedge}$), the $J_{\wedge}-J_x>0$ make the system favor stripe phase without topological defect while the $\delta \Sigma^x$ term favors more topological defects with flippable corners. The sector with many topological defects, which has much more freedom degree as the 'Sector III' of Fig.\ref{Fig2} (a) shown, is easy to be reached. Of course, we could set the 'Sector III' as the sector where the ground state in, but it seems trivial for an optimization problem because it is very easy to be found. Thus we set $\delta \ll J_{\wedge}-J_x$ in the following ($\delta=0$ for simplicity), i.e., the ground state is stripe as the 'Sector I' of Fig.\ref{Fig2} (a) shown. {Although the target state is a classical state, it doesn't lose the generality because the key point here is to jump among different topological sectors while the details of the ground state is not important.}
Our choice aims increasing the coefficient of difficulty of such optimization. 
%because the target state can also becomes a quantum state within other sector with certain number of topological defects via tunning the $\delta$ and $J$ terms~\cite{zhou2020quantum,ZZrk2021}. 
\begin{figure}[htp]
    \includegraphics[width=0.5\columnwidth]{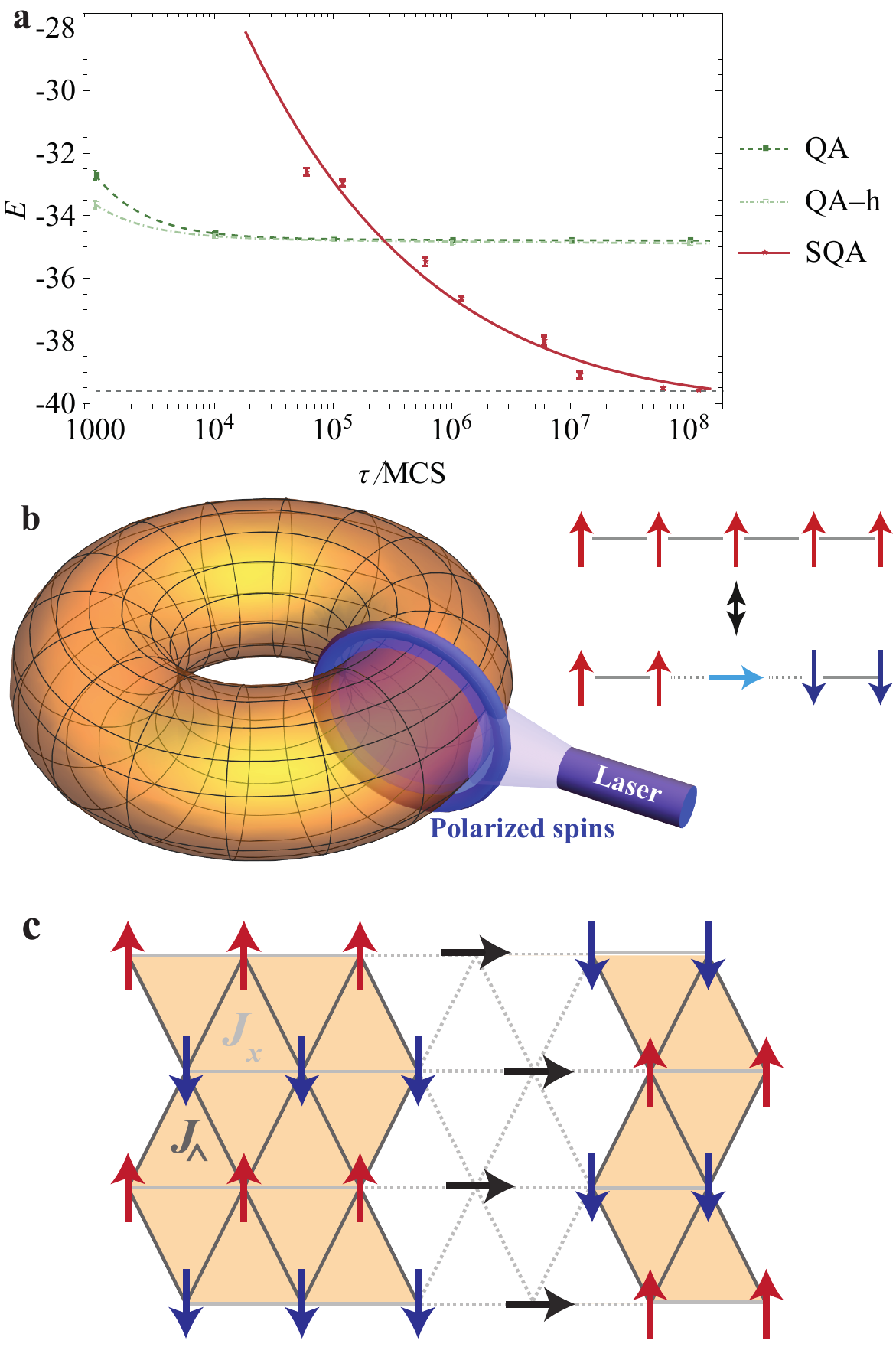}
    \caption{\textbf{The sweeping quantum annealing.} (a) The energy is computed for $6\times6$ system and the true ground state energy $E_g=-39.6J_{\wedge}$ is denoted by the grey dashed line. The expectation value of $E$ and its error bars are obtained from averaging over 64 independent annealing runs. The straightforward QA and QA-h have the worst annealing performance, the best among these is the SQA. In the whole paper, we use the statistical error to define the error bar, i.e., $\sigma/\sqrt{N_{sampling}}$, where the $\sigma$ is the standard error and $N_{sampling}$ is the number of QMC samplings. (b) Schematic diagram of SQA on a torus square lattice. We perform additional quantum annealing on `virtual edges' of the system to reduce the topological defects. In quantum simulation experiments, the global transverse field term can be controlled with tunable laser or arbitrary waveform generators~\cite{kim2010quantum,kim2011q,georgescu2014quantum}. (c) The SQA schematic diagram on a triangular lattice with the periodic boundary condition along both x and y directions. The polarized spins lose the Ising interactions which is similar to opening the boundary, as the white region shown.}
    \label{Fig3}
\end{figure}

\subsection{Sweeping quantum annealing scheme}
We first consider the conventional quantum annealing protocol, whose dynamics is generated by a Hamiltonian of the form
\begin{equation}
    H_{\textrm{QA}}=H_{\textrm{A}}-h(t)\sum_i\sigma^x_i,
    \label{QA}
\end{equation}
where $\sigma^x$ is the Pauli-x operator.
The annealing schedule is controlled by linearly reducing the transverse field $h(t)$ from a sufficiently large initial value $h(0)=5J_{\wedge}$ to the finial value $h(\tau)=0$, so that the original Ising Hamiltonian $H_{\textrm{A}}$ is recovered at the end.

In order to obtain the numerical results with large system size, we adopt the stochastic series expansion (SSE) method within continuous time frame~\cite{Sandvik1991,Sandvik1999,Sandvik2003,sandvik2019stochastic,Desai2021} to resolve the quantum dynamics, which essentially belongs to the quantum Monte Carlo (QMC) simulation~\cite{sm}. The QMC can reveal the scaling behaviors of QA as same as real time simulations and has been widely used to study the annealing problems~\cite{de2011universal,liu2013quasi,liu2014dynamic,liu2015quantum,Troyer2016,Neven2016,Neven2017,king2021scaling}.
We use the Monte Carlo step (MCS) to label the annealing time. One MCS is defined as visiting all the spins, so it is proportional to size $L^2$. However, as shown in Fig.~\ref{Fig3} (a), the QA protocol can not reach the ground state even for long enough evolution time.

We have also tried the quantum annealing with inhomogeneous transverse field, i.e., $H_{\textrm{QA-h}}=H_{\textrm{A}}+\sum_i h_i(t)\sigma^x_i$ with site-dependent random field $h_i(t)$, which are initially chosen from the uniform distribution $[0,10J_{\wedge}]$, and then linearly reduced to 0.
We denote such annealing as `QA-h', which is suggested to weaken the effect of second-order phase transition and improve the efficiency of quantum annealing~\cite{suzuki2007quantum,zurek2008phase,dziarmaga2010adiabatic,rams2016inhomogeneous,hauke2020perspectives}.
Nevertheless, as shown in Fig.~\ref{Fig3} (a), it behaves most like the conventional QA protocol and can not improve the efficiency.

%{\color{red}
The inefficiency of the above two QA protocols can be attributed to the following aspects. If the $\sigma^x$ breaks the `triangle rule', it will take an huge energy cost about $2J_{\wedge}$ which is forbidden at low temperature. Thus, the $\sigma^x$ operators always flip the spins at the corners of defects without Ising energy costs [Fig.~\ref{Fig2} (d)] and obtain extra kinetic energy advantage of the $\sigma^x$ term at the same time.
All these low-energy actions can not change the number of topological defects (i.e., the number of topological sectors), but just twist the corner by flipping the spin on it, such as the red spin of Fig.~\ref{Fig2} (d), after flipping the spin the corner on right side will be turned to left as shown by the dashed line. Therefore, the fluctuations introduced by both QA and QA-h can only deform the defects but cannot create/annihilate them. However, the topological optimization problems usually involve non-local excitations and require global operations.

In essence, the optimization problem here is an annealing problem to find the optimal topological sectors in Fig.~\ref{Fig2} (a) . In this case, the number of states in a topological sector increases with the defect number $N_{\rm D}$, so the stripe configuration of the ground state is the smallest sector and hard to be reached in Hilbert space as one needs an annealing algorithm that can both tunnel through different topological sectors and reduce the energy.
%from the following considerations: the local updates in TA and QA are hard to be accepted because of the breaking of the `triangular rule'; at the same time, the random field in QA-h didn't help in this regard but introduce too much disorder with more local minima. A proper global scheme of QA is to soften the constraint gradually to help the system to relax to a lower energy state and then gradually recover the constraint once the system is stabilized.
%This can be achieved via open boundary condition.
%}
\begin{figure}[htp]
    \includegraphics[width=0.6\columnwidth]{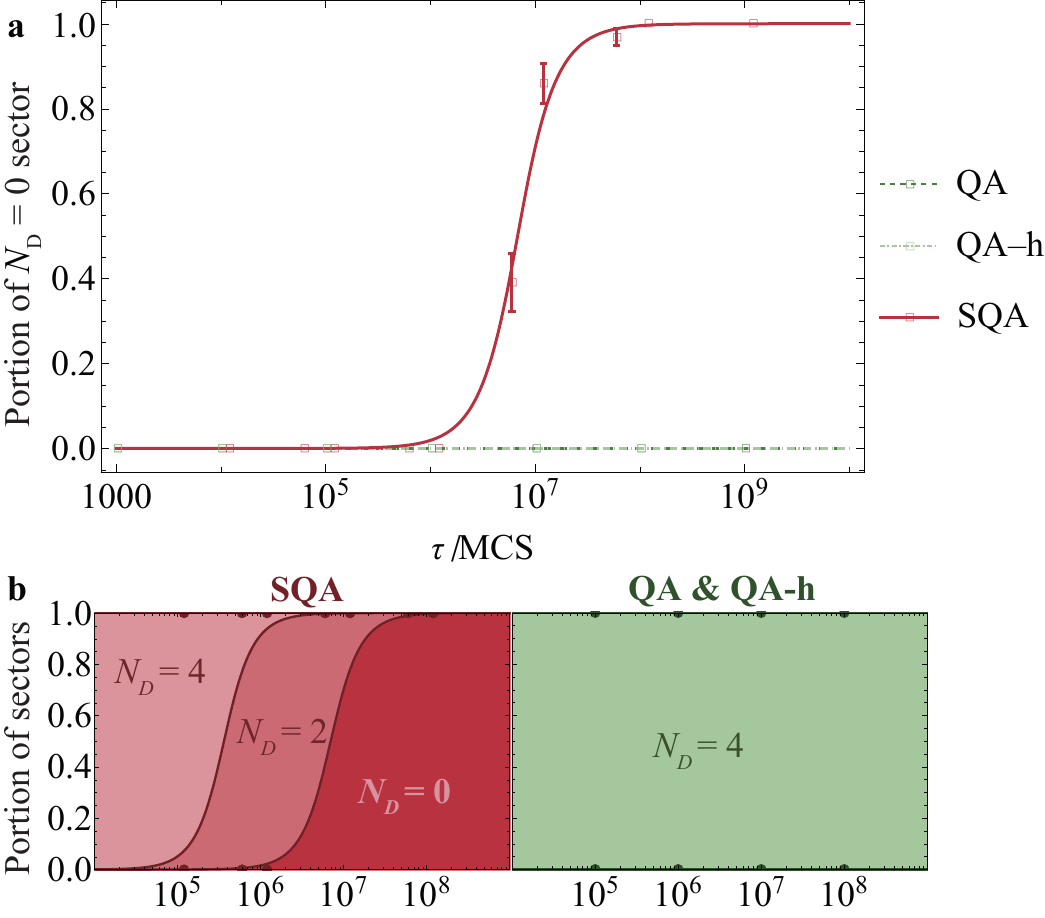}
    \caption{\textbf{The topological sector changes only by SQA.} (a) The proportion of ground state sector as a function of the annealing time with different methods. (b) The proportion of all topological sectors as a function of the annealing time with different methods.}\label{Fig4}
\end{figure}

In order to obtain non-local excitations to surmount the problem of the topological defects,
we develop the sweeping quantum annealing to achieve the ground state topological sector.
The key point is the topological defects can be moved out continuously through the 'open edge' by local operators. Thus the system can easily reach the global minimum intuitively if we can change the boundary condition during annealing.
The schematic diagram of SQA on a torus lattice taking square lattice as an example is shown as Fig.\ref{Fig3} (b). The feasible algorithmic scheme on an quantum annealing platform is to do additional annealing with a large transverse field on a virtual edge as the purple circle of Fig.\ref{Fig3} (b) shown. Intuitively, a large transverse field can polarize the spin to the $x$-axis, acting as if shearing the Ising coupling bond apart. As shown in the right part of Fig.\ref{Fig3} (b), the central spin has been polarized to reduce its Ising interaction with other spins. A great advantage is that the strength of the transverse field can be easily and continuously adjusted by tunning the Rabi frequency of laser in experiment. Similarly, the SQA on triangular lattice is shown in Fig.\ref{Fig3} (c), all the spins on the virtual edge are polarized, which is equal to opening the edge effectively.

We then scan all the virtual edges one by one in turn along a certain direction with strongly annealing transverse field, which is expected to achieve the effect across topological sectors. The specific plan is as follows: 1) Keep the conventional quantum annealing process on every site as mentioned above, that is, linearly decreasing the $h$ slowly down from $h(0)$ zero. 2) Divide the whole process into $L$ (system length) parts. In every part, add a extra strong transverse field (as same as $h(0)$) on virtual edges in order and reduce the field strength slowly to $h(t)$. $h(t)$ is the field strength of the traditional QA at the end of this part. Then move to the next virtual edge and repeat same process until all the edges ($1\sim L$) are visited.
From Fig.~\ref{Fig3} (a), we can see an obvious advantage of SQA to arrive at ground state quickly. The detailed pseudo-code protocol of SQA is shown in the Supplemental Note 2.
%In order to better understand the working mechanism of SQA, we give a physical explanation via a picture of flippable topological defects. Let's decompose the SQA into a sequence of cut-and-glue steps. In principle, we can cut the system to artificially create an open boundary~[Fig.~\ref{Fig3} (c)]. In order to allow the topological defect~\cite{zhang01,zhang02,zhang03} to pass through the boundary and disappear to change the topology, we need to make the spins at the open boundary flippable with nearly 0 energy cost. We change the coupling strengths of the bonds at the boundary to a different value $J_e$. The corresponding energy shift of one edge spin flipping process , e.g., the collapsed spin-up and spin-down of blue spin on edge in Fig.~\ref{Fig3} (c) is $2(2J_e-J_x)$. In order to soften the boundary, we set $J_e=J_x/2$ to make spin more flippable. After equilibrating the system with open boundary conditions for some time, the edges need to be glued to restore the original boundary conditions when utilizing the quantum annealing.

The physical reason of the superior behavior of SQA in the topological frustrated optimization problem is that it effectively removes topological defects in the systematic sweeping. As demonstrated in Fig.~\ref{Fig4} (a),
different from the  QA and QA-h, the proportion of ND = 0 sector are increasing with annealing time of SQA.
%the proportion of $N_{\rm D}=0$ sector are increasing with annealing time except for QA and QA-h}. %After $t/\textrm{MCS}\approx10^7$, the increasing rate of SQA surpasses the TA, and SQA exhibits the best optimization results.
Furthermore, the proportion of different topological sectors are shown in Fig.~\ref{Fig4} (b). The conventional QA and QA-h are stuck in $N_{\rm D}=4$ sectors, which reveals the reason for their inefficiency.
%Although the TA can also change the topological sectors due to the creation of pairs of vortex excitations in Fig.~\ref{Fig2} (f),
Clearly the best of all annealing schemes is the SQA which can straightforwardly change the topology of the system at any position. In order to strengthen the evidences of effectiveness and universality of SQA, we simulate another difficult topological optimization problem of a fully frustrated Ising system on square lattice (corresponding to square lattice quantum dimer model) in Supplementary Note 4, in where we see that our SQA protocol is also efficient.
\begin{figure}[htp]
    \includegraphics[width=0.85\columnwidth]{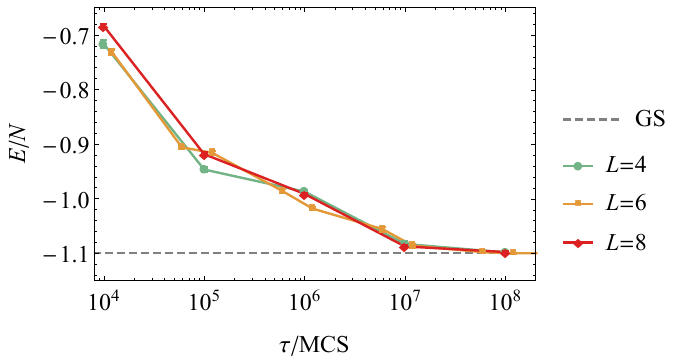}
    \caption{\textbf{Time complexity.} (a) The energy per site $E/N$ is computed for $4\times 4$, $6\times6$ and $8\times 8$ systems and the true ground state energy per site is denoted by the dashed line. The expectation value of $E/N$ and its error bars are obtained from averaging over 64 independent annealing runs. It shows a good convergence for all sizes. One MCS is defined as visiting all the spins, so MCS scales as $L^2$.}\label{Fig5}
\end{figure}

Remarkably, our SQA protocol can prepare the target topological state in a time that scales linearly with the system size ($N=L^2$). Fig.~\ref{Fig5} shows the energy per site $E/N$ as a function of MCS, in which every MCS is defined with scaling as $L^2$, i.e., $1 MCS \sim L^2$.
We see that the lines are almost coincide with each other for different system size, which clearly demonstrates a linear increase of annealing time with the system size, since the annealing time is proportional to MCS~\cite{Troyer2016,Neven2016,Neven2017,king2021scaling}. The power cost for the algorithm is a strong advantage.

{In addition, we have also discussed the effect of thermal annealing (TA)~\cite{Brooke1999,Brooke2001} in the Supplementary Note 6 although it is almost impossible realized in the recent cold atom experiments. Hopefully, the result may inspire related experiments in future.}

\subsection{Experimental candidates}
The experimental candidates for implementing the SQA protocol are many, for example in the superconducting circuits~\cite{wang2021observation,ge2021approximating,xu2020probing,Roushan21}.
And we find SQA is in particular applicable to the Rydberg arrays, which have recently been utilized to simulate the $\mathbb{Z}_2$ quantum spin liquid \cite{Semeghini21}.

Below we discuss in detail the Rydberg arrays systems are the idea platform to simulate the $U(1)$ LGT model [see Eq.~\eqref{HA}] and implement the SQA protocol.
The effective Hamiltonian of Rydberg arrays is~\cite{RevModPhys.82.2313, Browaeys2020NatPhys}
\begin{equation}
    H_{\rm R}=h \sum_{i} \sigma^x_i -\mu \sum_{i} n_i + V\sum_{i>j}\frac{n_i n_j}{|i-j|^6},
\label{HR}
\end{equation}
where $i$ and $j$ are the site labels, $n_i=0,~1$ is the density operator to probe the ground state or Rydberg state, respectively, and $\sigma^x$ is the tunneling term (Pauli-x operator) to connect the two states.
We can use the spin-1/2 language to describe the qubit within the two-level Rydberg system, since there is a common mapping between the hardcore boson and spin, that is, $n =(\sigma^z+1)/2$ and ground/Rydberg state corresponds to spin down/up state.
%The above Hamiltonian [Eq.~\eqref{HR}] can then be recast into the form of the $U(1)$ lattice gauge model [Eq.~\eqref{HA}], if we choose a triangular array and certain suitable parameters $h,\,\mu,\, V$, such as the prediction of Ref.\cite{ZZrk2021}.

The Rydberg arrays Hamiltonian [Eq.(\ref{HR})] on triangular lattice can then be cast into the form of [Eq.~\eqref{QA}] with emergent $U(1)$ LGT, it has been realized in experiment in fact~\cite{Rydberg_nature2}. Firstly, since the repulsive interaction decays very fast, i.e. $1/r^6$, the strength of second nearest neighbours on triangular lattice is about 0.037 compared with that of the first neighbour, so the model can be further simplify as $H_{\rm R}=h \sum_{i} \sigma^x_i -\mu \sum_{i} n_i + V\sum_{\langle ij\rangle}{n_i n_j}$ with only nearest neighbours. Then, we set $\mu=V$ and use $n_i=2\sigma^z_i+1$ to simplify the Hamiltonian in $\sigma$ operator form as $H_R=h \sum_{i} \sigma^x_i +\frac{V}{4} \sum_{\langle ij\rangle} \sigma^z_i \sigma^z_j.$ Lastly, set the distances of Rydberg atoms along x-axis a little farther than along other directions, and then the anisotropy Hamiltonian in Eq.~\eqref{QA}, $H_R=h \sum_{i} \sigma^x_i +\frac{V_{\wedge}}{4} \sum_{\langle ij\rangle_{\wedge}} \sigma^z_i \sigma^z_j+\frac{V_x}{4} \sum_{\langle ij\rangle_x} \sigma^z_i \sigma^z_j$, is obtained. In this way, the superior behavior of SQA we have discussed for Eq.~\eqref{QA}, can be readily applied to the Rydberg arrays. Since all these parameters can be tuned by adjusting laser detuning or Rabi frequency, we think the Rydberg arrays is the a well-suited platform to implement the SQA protocol.

\section{Discussion}
As the fast development of Rydberg arrays simulation, precise regulation of the quantum state of LGT within topological sector is required. For such topological optimization problems, there is still a lack of systematic quantum simulation protocol. In this work, we find that the emergent topological properties in frustrated Ising systems greatly reduce the efficiency of both conventional thermal and quantum annealing. Borrowing the idea of changing topology by cutting and gluing the system, we invent such a generalized algorithm --- the sweeping quantum annealing method to solve these problems with huge quantum speedup. Moreover, the SQA can be easily implemented in realistic quantum simulation experimental platforms because the global transverse field term can be controlled finely by tunable laser or arbitrary waveform generators~\cite{kim2010quantum,kim2011q,georgescu2014quantum,EE2010PRB,king2018observation,ES2020PRL,wang2021observation,ge2021approximating,xu2020probing}. After comparing with conventional quantum
and sweeping quantum annealing algorithm, we find that the SQA presents high efficiency and validity while other annealing schemes fail. The sweeping quantum annealing therefore opens up an effective and innovative way for controlling quantum states of LGT. It will also be of interests to study the effectiveness of SQA under noises and dissipations~\cite{ZC2014,ZY2018} which commonly exist in the quantum simulator.

It's worth noting that the topological defects are along one direction in this work. Topological defects along the other direction can also be removed under the SQA along that direction. Therefore, the system with defects in both directions can be optimized by SQA mixed in both directions in principle. In the future works, more general systems with both nearly degenerate topological sectors in two directions and glassy minimums will be discussed.

\section{Method}
\textbf{Quantum Monte Carlo simulation:}
We use continue time quantum Monte Carlo method for the quantum annealing in this paper. Because the discrete time QMC may take opposite result due to the error~\cite{Troyer2016}. The temperature $T=1/\beta$ we set is $J_{\wedge}/20$, which is much smaller than the energy scale of spinon excitation $\sim 2J_{\wedge}$. The temperature can keep the configurations obeying the `triangle rule' with emergent LGT. The dynamic of annealing can be simulated via QMC because the time scaling of QMC is same as real time evolution, these simulations have been widely used to study the quantum annealing behaviours~\cite{Troyer2016,Neven2016,Neven2017,king2021scaling}.

The QMC here we used is stochastic series expansion (SSE) algorithm~\cite{Sandvik1991, Sandvik1999, Sandvik2003}. The partition function $Z$ is dealt with a Taylor expansion as:
\begin{equation}
    Z=\sum_\alpha\sum_{n=0}^\infty\frac{\beta^n}{n!}\langle\alpha|(-H)^n|\alpha\rangle
\end{equation}
Then we extract the wanted information via sampling these configurations of Taylor expansion.
More details of quantum Monte Carlo simulations for QA, QA-h and SQA have been explained in the Supplementary Note 1.

\section{Acknowledgments}
We especially acknowledge Xiaopeng Li for useful discussions and suggestions. We also wish to thank X. X. Yi, Heng Fan, Z. Y. Ge and Shangqiang Ning for constructive discussions. ZY thanks the start-up fund of Westlake University. 
ZY and ZYM acknowledges the support from the Research Grants Council of Hong Kong SAR of China (Grant Nos. 17301420, 17301721, AoE/P-701/20, 17309822, HKU C7037-22G), the ANR/RGC Joint Research Scheme sponsored by Research Grants Council of Hong Kong SAR of China and French National Reserach Agency(Porject No. A\_HKU703/22) and the Seed Funding “Quantum-Inspired explainable-AI” at the HKU-TCL Joint Research Centre for Artificial Intelligence. 
X.-F. Z. acknowledges funding from the National Science Foundation of China under Grants  No. 12274046, No. 11874094 and No.12147102, Chongqing Natural Science Foundation under Grants No. CSTB2022NSCQ-JQX0018, Fundamental Research Funds for the Central Universities Grant No. 2021CDJZYJH-003. The authors acknowledge Beijng PARATERA Tech Co.,Ltd.(\url{https://www.paratera.com/}) for providing HPC resources that have contributed to the research results reported within this paper. Y.C.W. acknowledges  support from Zhejiang Provincial Natural Science Foundation of China (Grant Nos. LZ23A040003). X. Q. acknowledges support from National Natural Science Foundation of China (Grants No. 12104098). Z.Z. acknowledges supports from the Natural Sciences and Engineering Research Council of Canada (NSERC) through Discovery Grants.

\section*{DATA AVAILABILITY}
The data that support the findings of this study are available from the authors upon reasonable request.

\section*{CODE AVAILABILITY}
The code is available from the authors upon reasonable request

\section{COMPETING INTERESTS}
The authors declare no Competing Financial or Non-Financial Interests.

\section{AUTHOR CONTRIBUTIONS}
Z.Y. and Z.Z. contributed equally to this work (co-first author). Z.Y., Z.Y.M. and X.F.Z. initiated the work. Z.Y. put forward the SQA scheme. Z.Y. and Z.Z. performed the QMC computational simulations. Y.H.Z. did the real time evolution in small size. All authors contributed to the analysis of the results. X.Q., Z.Y.M. and X.F.Z. supervised the project.

%\bibliography{QA}

\end{document}

\title{Supplemental Information (Quantum optimization within lattice gauge theory model on a quantum simulator)}
\author{Zheng Yan}%\textsuperscript{1}}
%\thanks{These authors contributed equally to this work.}
%\email{zhengyan@westlake.edu.cn}
%\affiliation{Department of Physics, School of Science, Westlake University, 600 Dunyu Road, Hangzhou 310030, Zhejiang Province, China}
%\affiliation{Institute of Natural Sciences, Westlake Institute for Advanced Study, 18 Shilongshan Road, Hangzhou 310024, Zhejiang Province, China}

\author{Zheng Zhou}%\textsuperscript{1}}
%\thanks{These authors contributed equally to this work.}
%\affiliation{State Key Laboratory of Surface Physics and Department of Physics, Fudan University, Shanghai 200438, China}
%\affiliation{Perimeter Institute for Theoretical Physics, Waterloo, Ontario N2L 2Y5, Canada}
%\affiliation{Department of Physics and Astronomy, University of Waterloo, Waterloo, Ontario, Canada N2L 3G1}

\author{Yan-Hua Zhou}
%\affiliation{Department of Physics, and Center of Quantum Materials and Devices, Chongqing University, Chongqing 401331, China}
%\affiliation{Chongqing Key Laboratory for Strongly Coupled Physics, Chongqing University, Chongqing 401331, China}

\author{Yan-Cheng Wang}
%\affiliation{Zhongfa Aviation Institute of Beihang University, Hangzhou 310023, China}
%\affiliation{Tianmushan Laboratory, Hangzhou 310023, China}

\author{Xingze Qiu}
%\email{xingze@fudan.edu.cn}
%\affiliation{State Key Laboratory of Surface Physics and Department of Physics, Fudan University, Shanghai 200438, China}

\author{Zi Yang Meng}
%\email{zymeng@hku.hk}
%\affiliation{Department of Physics and HKU-UCAS Joint Institute of Theoretical and Computational Physics, The University of Hong Kong, Pokfulam Road, Hong Kong, China}
%\affiliation{Beijing National Laboratory for Condensed Matter Physics and Institute of Physics, Chinese Academy of Sciences, Beijing 100190, China}
%\affiliation{Songshan Lake Materials Laboratory, Dongguan, Guangdong 523808, China}

\author{Xue-Feng Zhang}
%\email{zhangxf@cqu.edu.cn}
%\affiliation{Department of Physics, and Center of Quantum Materials and Devices, Chongqing University, Chongqing 401331, China}
%\affiliation{Chongqing Key Laboratory for Strongly Coupled Physics, Chongqing University, Chongqing 401331, China}

%\renewcommand{\thefootnote}{*}
%\footnotetext{These authors contributed equally to this work.}
\maketitle
%\tableofcontents

%\section{Classical Monte Carlo for thermal annealing}
%
%We use Metropolis Monte Carlo~\cite{Metropolis1953} which flip single spin to simulate Eq.~(\ref{HA}) thermal annealing. Its general process is as follows: assume we have reached configuration $C_1$ with energy $E_1$. Firstly we choose a spin randomly and flip it to get a new configuration $C_2$ with energy $E_2$. We use the acceptance probability $P=\min(e^{-\beta(E_2-E_1)},1)$ to determine whether to update the system to configuration $C_2$, otherwise we keep configuration $C_1$. For the thermal annealing process, we do the Monte Carlo simulation from $T=5$ to $T=0.05$, the decreasing interval is $\Delta T=0.05$, and MCS of every interval is fixed.

\section{Supplementary Note 1}
\textbf{Quantum Monte Carlo for quantum annealing.}
For the quantum annealing in this paper, we use a quantum Monte Carlo (QMC) method with stochastic series expansion (SSE) algorithm\cite{Sandvik1991, Sandvik1999, Sandvik2003}. In this method, the evaluation of partition function $Z$ is done by a Taylor expansion, and the trace is taken by summing over a complete set of suitably chosen basis.
\begin{equation}
    Z=\sum_\alpha\sum_{n=0}^\infty\frac{\beta^n}{n!}\langle\alpha|(-H)^n|\alpha\rangle
\end{equation}
By writing the Hamiltonian as the sum of a set of operators whose matrix elements are easy to calculate $H=-\sum_iH_i$ and truncating the Taylor expansion at a sufficiently large cutoff $M$, we can further obtain
\begin{equation}
    Z=\sum_\alpha\sum_{\{i_p\}}\beta^n\frac{(M-n)!}{M!}\langle\alpha|\prod_{p=1}^nH_{i_p}|\alpha\rangle
\end{equation}
To carry out the summation, a Markov chain Monte Carlo procedure can be used to sample the operator sequence $\{i_p\}$ and the trial state $\alpha$. One step of the update process contains diagonal update and off-diagonal update. In the diagonal update, the diagonal operators are inserted into and removed from the operator sequence. Meanwhile, in the off-diagonal update, the diagonal and off-diagonal operators can be converted into each other. In the quantum annealing process, we decrease the transverse field from $h=5.00$ to $h=0.00$, with a decreasing interval $\Delta h=0.05$. A tiny residual field is kept to facilitate the update process.

\section{Supplementary Note 2}
\textbf{Pseudo-code of SQA.}
The details of sweeping quantum annealing (SQA) algorithm can be written via a pseudo code as Table.\ref{Tbl1}.
\begin{table}[htb]
    \caption{The pseudo-code demonstrating the sweeping annealing procedure.}
    \label{Tbl1}
    \begin{tabular}{p{.05\linewidth}p{.4\linewidth}p{.5\linewidth}}
        \hline\hline
        1.&\textbf{Define} steps&\cm{annealing steps in each $\Delta h$ window.}\\
        2.&\textbf{Define} $n=L$&\cm{total \# of cutting position.}\\
        3.&\textbf{Set} $n_s=\mathrm{steps}/n$&\cm{\# of annealing steps for gluing.}\\
        4.&\textbf{For} $h=h_\mathrm{max}:-\Delta h:0$&\cm{annealing loops}\\
        5.&\quad \textbf{For} $i_s=1:n$&\cm{sweeping loops}\\
        &\multicolumn{2}{l}{\qquad\cm{----------Cutting Process----------}}\\
        6.&\multicolumn{2}{l}{\qquad\textbf{Set} $h_e=h_{max}$}\\
        &\multicolumn{2}{l}{\qquad\cm{$h_e$ denote h acts on the edge.}}\\
        &\multicolumn{2}{l}{\qquad\cm{----------Gluing Process----------}}\\
        7.&\multicolumn{2}{l}{\qquad\textbf{Set} $\Delta h_e=(h_{max}-h)/n_s$}\\
        &\multicolumn{2}{l}{\qquad\cm{the increment of gluing.}}\\
        8.&\qquad\textbf{For} $i_g=1:n_s$ &\cm{gluing loops}\\
        9.&\multicolumn{2}{l}{\qquad\quad\textbf{One annealing step with QMC}}\\
        10.&\multicolumn{2}{l}{\qquad\quad\textbf{Set} $h_e=h_e-\Delta h_e$}\\
        &\multicolumn{2}{l}{\qquad\quad\cm{smoothly restoring the boundary condition}}\\
        11.&\qquad\textbf{End}&\\
        12.&\multicolumn{2}{l}{\qquad\textbf{Move to next cutting position}}\\
        13.&\quad\textbf{End}&\\
        14.&\textbf{End}&\\
        \hline
    \end{tabular}
\end{table}~~~~~
\begin{figure*}[htp]
    \includegraphics[width=0.8\linewidth]{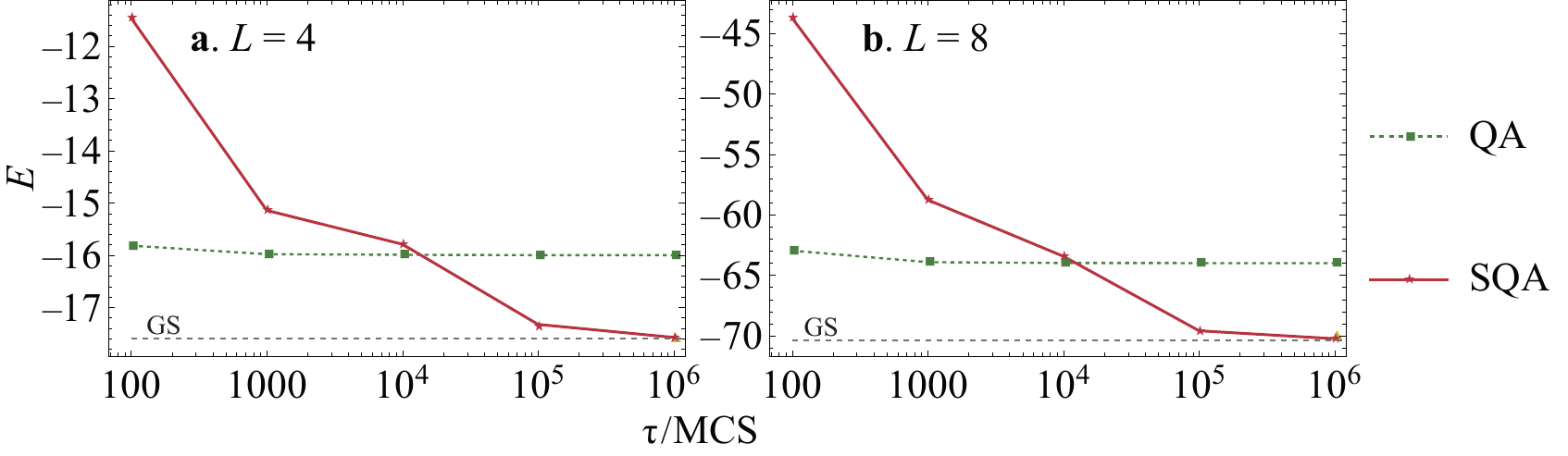}
    \caption{The energy is computed for $4\times 4$ and $8\times 8$ system and the true ground state energy is denoted by the grey dashed line.}
    \label{FigS1}
\end{figure*}

The "steps" of the pseudo-code is the number of annealing steps in each $\Delta h$ window. Notice that the $h_{max}$ and $-\Delta h$ are fixed, so that the total Monte Carlo steps is equal to $steps*h_{max}/\Delta h$ which is also unchanged. (It also explains the time cost of Fig.5 of main text that, as we set a total MCS number to process all the annealing actions, when the L becomes larger, each edge annealing will obtain less time. Therefore, the total time cost is just related to the cost of every MCS.) The $n$ is equal to $L$ which is the total number of cutting position, that means, all the virtual edges should be opened once time. The annealing loops means we divide the whole annealing process into $h_{max}/\Delta h$ pieces. Because every piece has similar procedure, we call them as ``annealing loops". In each annealing loop, we have the "sweeping loops". what the sweeping loops do is opening the virtual edge and gluing it slowly. Thus we have set $h_e=h_{max}$ initially and act the transverse field $h_e$ on the virtual edge spins. Then, the $\Delta H_e$ is also set for the gluing process. In each gluing loop, the $h_e=h_e-\Delta h_e$. Through quantum Monte Carlo simulation, the virtual edge glues until $h_e=h$. Then we do same process on the next virtual edge, until all the edges are visited.

\section{Supplementary Note 3}
\textbf{Annealing results of different sizes.}
We also simulate QA and SQA on $4 \times 4$ and $8 \times 8$ triangular lattice to see the corresponding annealing effects. As can be seen from the Fig.\ref{FigS1}[(a) for 4x4, (b) for 8x8], the leadership of SQA does not change because of the lattice size. Besides, we can also intuitively feel that larger size needs longer annealing time. Moreover, the superiority of SQA becomes more obvious in larger size.
%\begin{figure}[htp]
%    \includegraphics[width=0.7\linewidth]{square.pdf}
%    \caption{(a) A fully frustrated Ising model on square lattice. The thick gray bonds are antiferromagnetic, and the thin black bonds are ferromagnetic. The spins of one bond is not low energy that creates a dimer on its dual bond. The low energy fully frustrated rule is equal to the condition that there must be and only be one dimer on every site of dual lattice. We set the links crossed by dimers a little weaker to let dimers condensed on the corresponding positions. Therefore the staggered sector becomes the lowest energy one. (b) The energy is computed for $6\times 6$ system and the true ground state energy is denoted by the grey dashed line.}
%    \label{square}
%\end{figure}
\begin{figure}[htp]
    \includegraphics[width=0.85\linewidth]{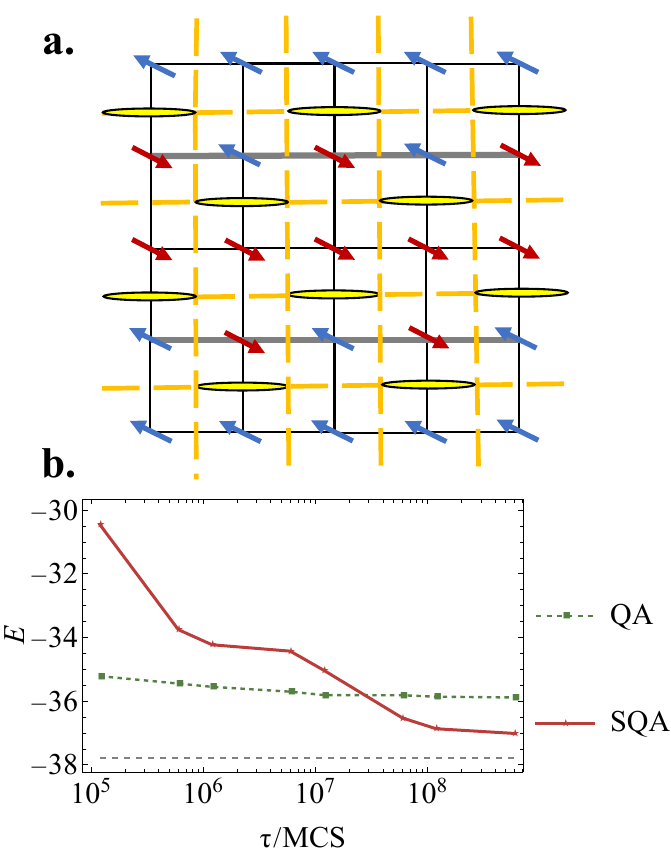}
    \caption{(a) A fully frustrated Ising model on square lattice. The thick gray bonds are antiferromagnetic, and the thin black bonds are ferromagnetic. The spins of one bond is not low energy that creates a dimer on its dual bond. The low energy fully frustrated rule is equal to the condition that there must be and only be one dimer on every site of dual lattice. We set the links crossed by dimers a little weaker to let dimers condensed on the corresponding positions. Therefore the staggered sector becomes the lowest energy one. (b) The energy is computed for $6\times 6$ system and the true ground state energy is denoted by the grey dashed line.}
    \label{FigS2}
\end{figure}

\section{Supplementary Note 4}
\textbf{Fully frustrated Ising model on square lattice.}
Similar to the frustrated Ising model on triangular lattice in main text, we give a fully frustrated Ising model on square lattice. The Hamiltonian to be annealed is as following:
\begin{equation}
    H_{\textrm{A}}=\sum_{\langle ij\rangle}J_{ij}\sigma^z_i\sigma^z_j,
\label{HA}
\end{equation}
i,j here mean the positions of spins of nearest neighbors. As in Fig.\ref{FigS2} (a), the thick gray bonds are antiferromagnetic (AFM) (set $J_{ij}=1$), and the thin black bonds are ferromagnetic (FM)(set $J_{ij}=-1$). What's more, we set the links crossed by dimers a little weaker ($J_{ij}=0.9$ for AFM and -0.9 for FM) to let dimers condensed on the corresponding positions. The low energy fully frustrated rule is equal to the condition that there must be and only be one dimer on every site of dual lattice, which contains an emergent gauge field constraint. The anisotropy make the staggered sector become the lowest energy one. Therefore, the model has similar topology annealing problem as in Fig.1 of main text.

We compare QA and SQA similarly as in main text. The SQA is also the best scheme as Fig.\ref{FigS2} (b) shown.

\section{Supplementary Note 5}
\textbf{Real-time evolution for the QA and SQA.}
To see the effectiveness of SQA on real-time evolution, we compare the time-evolution of SQA and QA for a $3\times2$ system using the Runge-Kutta algorithm. 
\par
\begin{figure}[htp]
    \includegraphics[width=0.85\linewidth]{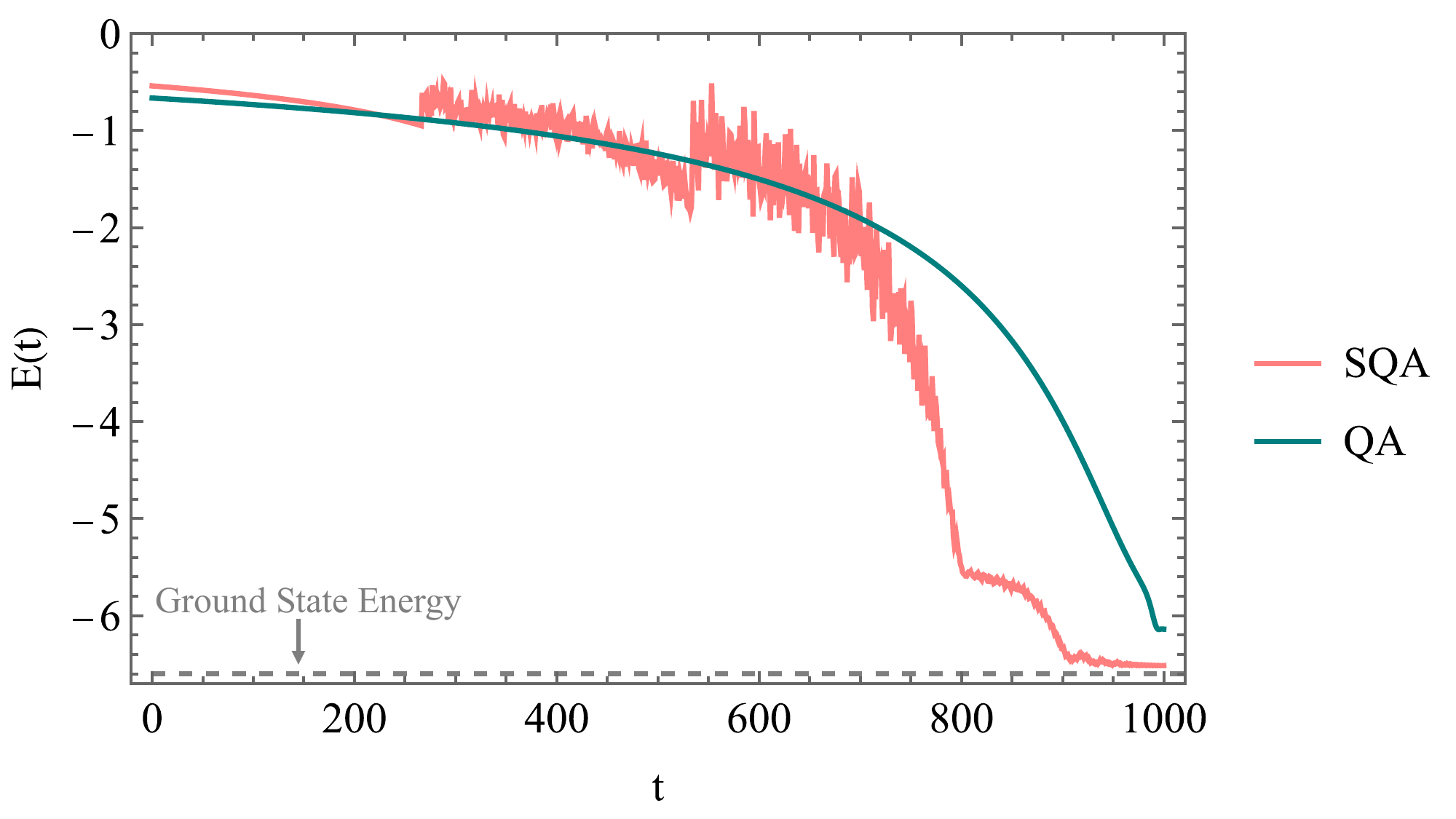}
    \caption{\label{compare_RK}The relation between the energy of $H_A$ and real time. The real-time evolution happens on a $3\times 2$ triangular lattice .}
    \label{compare}
\end{figure}

The initial state $\psi(0)$ is the eigenstate of the Hamiltonian at $t=0$ and we measure the energy $E(t)=\bra{\psi(t)}H_A\ket{\psi(t)}$. The result is shown in Fig. \ref{compare_RK}. As shown in this figure, the quantum state reaches a high fidelity under SQA evolution while using less time, even though the length of topological defect is about $L_y=2$ in the small size and the topological defect is easier to be overcome.
The results has proven that SQA is more effective compared with QA. 
\begin{figure}[htp]
    \includegraphics[width=0.85\linewidth]{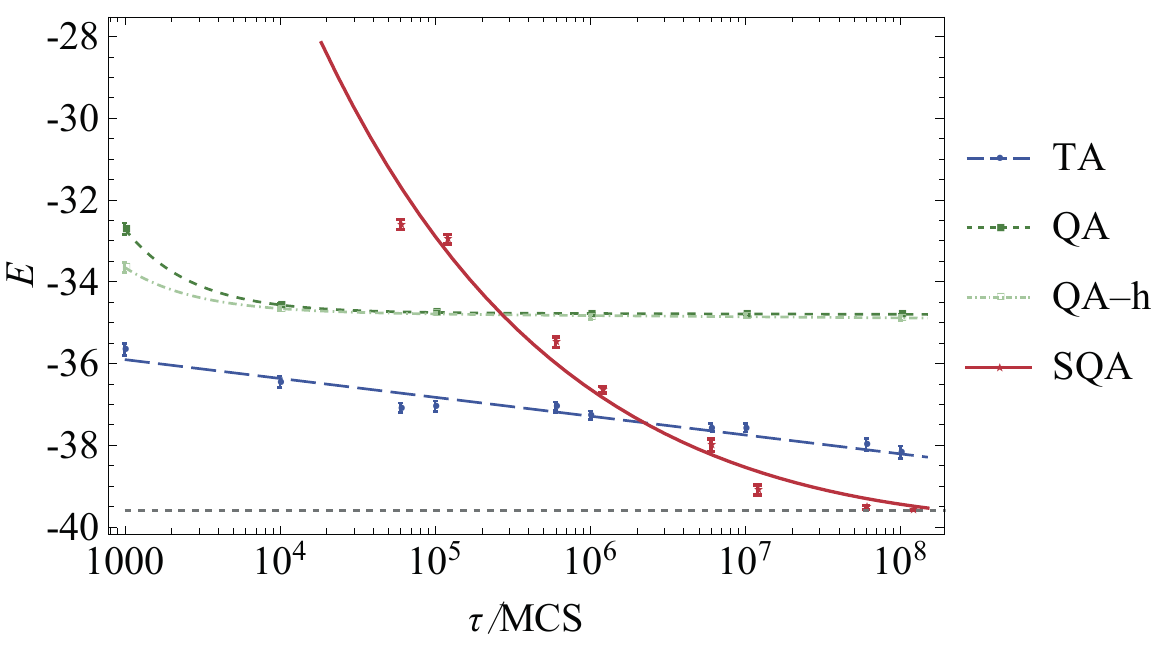}
    \caption{The energy is computed for $6\times6$ triangular system and the true ground state energy $E_g=-39.6J_{\wedge}$ is denoted by the grey dashed line. The expectation value of $E$ and its error bars are obtained from averaging over 64 independent annealing runs. The straightforward QA and QA-h have the worst annealing performance and the TA works better, the best among these is the SQA. In the whole paper, we use the statistical error to define the error bar, i.e., $\sigma/\sqrt{N_{sampling}}$, where the $\sigma$ is the standard error and $N_{sampling}$ is the number of QMC samplings. }
    \label{FigS3}
\end{figure}

\section{Supplementary Note 6}
\textbf{The thermal annealing for this problem.}
Although the thermal annealing (TA) is not available in the cold atom experiment so far, we have also added its result in Fig. \ref{FigS3} for the reference of the experiment in future.
In the thermal annealing, we set the Boltzmann constant $k_\mathrm{B} = 1$ for convenience. By comparing thermal annealing and quantum annealing, study how the energies $E$ evolute under same annealing time (MCS). We do thermal annealing through deceasing linearly from $T=5$ to $T=0.05$, cooling interval is $\Delta T= 0.05$. We parallelize 64 sets of independent Markov chain calculations and average the results to get the final data.

It's obvious that the TA is better than QA but much slower than SQA. The reason is that the spinons (triangle with 3 parallel spins) connected with defects as shown in Fig.2 (f) of mian text are easy to be excited at higher temperature, so that the topological sector can be changed in TA. But while the temperature is slowly cooling down, the spinons become harder to be excited which makes the effectiveness of TA is not as good as SQA.

%\bibliography{SM}